\documentclass{amsart}
\usepackage{amssymb, latexsym}
\begin{document} 
\title{\LARGE \textbf A Lorentz Covariant Noncommutative Geometry}
\author{A. Lewis Licht}
\address{Dept. of Physics\\University of Illinois at Chicago\\Chicago, 
Illinois 60607\\licht@uic.edu}

\begin{abstract}
A noncommutative geometry that preserves lorentz covariance was introduced 
by Hartland Snyder in 1947.  We show that this geometry has unusual properties 
under momentum translation, and derive for it a form of star product. 
This work was presented at the 7th International Wigner Symposium, 
August 2001, College Park, MD, USA.
\end{abstract}
\maketitle
\section{Introduction}
The possibility that space time may be noncommutative in the sense that:
\begin{equation}
[x^{\mu}, x^{\nu}] = i\theta^{\mu \nu}
\end{equation}
where $\theta^{\mu \nu}$ is a constant, has appeared recently in string
theory,~\cite{douglas} and also in measurement theory.~\cite{sak}  The
$\theta^{\mu \nu}$ may be thought of as a kind of background field, 
and its presence implies some small violation of Lorentz invariance.  
The possibility of detecting this Lorentz violation has been discussed 
in Ref.~\cite{carroll}.

It follows from Eq.(1) that:
\begin{equation}
\Delta x^{\mu} \Delta x^{\nu} \geq | \theta^{\mu \nu} |
\end{equation}
and  $\theta^{\mu \nu}$ provides some sort of minimal length.  There are
measurement theory arguments that suggest that  $\theta^{\mu \nu}$ should
be of the order of the square of the Planck length 
$\ell^{2}_{P}$.~\cite{sak}

In 1947, Hartland S. Snyder~\cite{snyd} suggested that a minimal length might be a
natural way of introducing a high momentum cut-off into field theory 
calculations.  To this end, he introduced a Lorentz covariant noncommutativity, 
which, in modern notation, takes the form:   
\begin{equation}
[x^{\mu}, x^{\nu}] = i\theta M^{\mu \nu}
\end{equation}
where here $\theta$ is a constant, and the $M^{\mu \nu}$ are the 
generators of the Lorentz transformations, satisfying:
\begin{equation}
[M^{\mu \nu}, M^{\rho \sigma}] = i[\eta^{\mu \rho}M^{\nu \sigma} - \eta^{\mu 
\sigma}M^{\nu \rho} + \eta^{\nu \rho}M^{\mu \sigma} - \eta^{\nu \sigma}M^{\mu 
\rho}]
\end{equation}
\begin{equation}
[M^{\mu \nu}, x^{\rho}] = i[\eta^{\mu \rho}x^{\nu} - \eta^{\nu \rho}x^{\mu}] \nonumber
\end{equation}
\begin{equation}
[M^{\mu \nu}, P^{\rho}] = i[\eta^{\mu \rho}P^{\nu} - \eta^{\nu 
\rho}P^{\mu}]\nonumber
\end{equation}
where $\eta^{\mu \nu}$ is the metric tensor, = (- , + , + , + ) , and 
$P^{\rho}$ is the momentum 4-vector, assumed to satisfy:  
\begin{equation}
[P^{\mu}, P^{\nu}] = 0
\end{equation}
and, to be consistent with the Jacobi identity, the canonical x-P commutation relation must be replaced by:
\begin{equation}
[x^{\mu}, P^{\nu}] = i[\eta^{\mu \nu} + \theta P^{\mu}P^{\nu}]
\end{equation}

\section{Snyders Representations}
Snyder found two representations for this space.  The first representation works 
for positive $\theta$.  It was based on a 5 dimensional Minkowski manifold, 
$\{\eta^{m}, m = 0,..4\}$, with metric $g_{\mu \nu} = \eta_{\mu \nu}$, 
$g_{\mu 4} = 0$, $\mu ,\nu = 0,...3, g_{44} = +1$.  The coordinate 
operators $x^{\mu}$ he identified with certain of the generators $M^{mn}$, m,n = 0,..4, of SO(1,5), 
the Lorentz group acting on this manifold:
\begin{equation}
x^{\mu} = aM^{\mu 4}
\end{equation}
where $a^{2} = \theta$.

For the momentum operators he found:
\begin{equation}
P^{\mu} = \frac{\eta^{\mu}}{a\eta^{4}}
\end{equation}
In this representation, it can be seen that the $x^{i}$ and the $M^{ij}$, 
with i, j = 1,2,3, are rotation operators for the four space-like coordinates, 
thus they generate SO(4), and it follows that the  $x^{i}$ have discrete eigenvalues that are multiples of a.
The time coordinate, $x^{0}$, is a boost operator in the $\eta^{4}$ direction, and therefore has continuous 
eigenvalues.

Snyder's second representation works for both positive and negative $\theta$.  
In modern notation, it is based on ordinary canonical 4-vector position 
operators $\xi^{\mu}$ , obeying the commutation relations:
\begin{equation}
[\xi^{\mu}, \xi^{\nu}] = 0
\end{equation}
\begin{equation}
[\xi^{\mu}, P^{\nu}] = i\eta^{\mu \nu} \nonumber
\end{equation}
and, defining a dilation operator as:
\begin{equation}
D = \frac{1}{2}\{\xi^{\alpha}, P_{\alpha}\}
\end{equation}
then Snyder's coordinate operators can be written as:
\begin{equation}
x^{\alpha} = \xi^{\alpha} + \frac{\theta}{2}\{P^{\alpha}, D\}
\end{equation}

\section{Translations in Momentum Space}

In the following, we will use a dot to indicate the 4-vector inner product, 
and hatted symbols to denote operators.  The operators
\begin{equation}
U_{0}(k) = e^{ik\cdot \hat{\xi}}
\end{equation}
perform translations in momentum space, 
\begin{equation}
U_{0}(k)|q^{\alpha}> = |q^{\alpha} + k^{\alpha}>
\end{equation}
where  $| q >$ is an eigenket of the 4-momentum operator.  The corresponding operators 
in the Snyder geometry:
\begin{equation}
U(k) = e^{ik\cdot \hat{x}}
\end{equation}
do not act as simply.  On a momentum eigenket they give:
\begin{equation}
U(k)|q^{\alpha}> = A|q^{\alpha}g + k^{\alpha}f>
\end{equation}
where A, g and f are functions of $k^{2}$ and $k\cdot q$:
\begin{eqnarray}
g = \left[ cos(\phi) - k\cdot q\sqrt{\frac{\theta}{k^{2}}}sin(\phi)\right]^{-1}\\
f = \left[ k\cdot q\left( cos(\phi) - 1\right ) + 
\sqrt{\frac{k^{2}}{\theta}}sin(\phi)\right ] g/k^{2}\\
A = \left |det \frac{\partial(q^{\alpha}g + k^{\alpha}f)}{\partial 
q^{\beta}} \right |^{1/2}
\end{eqnarray}
here
\begin{equation}
\phi = \sqrt{\theta k^{2}}
\end{equation}
To first order in $\theta$ , these become:
\begin{eqnarray}
g = 1 + \theta \left (\frac{k^{2}}{2} +  k\cdot q \right )\\
f = 1 + \theta \left (\frac{k^{2}}{3} +  \frac{k\cdot q}{2} \right )\\
A = 1 + \frac{5}{4}\theta \left ( k^{2} +  2k\cdot q \right )
\end{eqnarray}

Let us denote the transformed momentum, $q^{\alpha}g + k^{\alpha}f$ , by 
$\acute{q}^{\alpha}$ .  The transformation has the strange property that $\acute{q}^{\alpha}$ 
becomes infinite for those momenta $k^{\alpha}$ that satisfy:
\begin{equation}
cot \left(\sqrt{\theta k^{2}} \right) = k\cdot q\sqrt{\frac{\theta}{k^{2}}}
\end{equation}
This is simplest when the initial momentum $q^{\alpha}$ is zero.  Then $\acute{q}^{\alpha}$ 
becomes infinite when
\begin{equation}
k^{2} = \frac{(n\pi)^{2}}{\theta}
\end{equation}
for any integer n.  If we take $\theta$ as positive, these are spacelike momenta, and 
a negative $\theta$ would give us infinities at timelike k.  The measurement 
theory arguments, with $\theta$ proportional to $\ell^{2}_{P}$ , 
imply that these are multiples of the Planck mass.

\section{The Star Product}

For operators that satisfy Eq. (1) with a constant matrix, there is a one to one correspondence 
between functions f on Minkowski space and operators F on Hilbert space, 
$F \Longleftrightarrow f$, given by:
\begin{equation}
F(\hat x) = \int \frac{d^{4}k}{(2\pi)^2}e^{ik \cdot \hat{x}}\tilde{f}(k)
\end{equation}
where $\tilde{f}$ is related to a function f in position space:
\begin{equation}
\tilde{f}(k) = \int \frac{d^{4}x}{(2\pi)^2}e^{-ik \cdot x)}f(x)
\end{equation}
and also can be related to the operator F:
\begin{equation}
\tilde{f}(k) = Tr\left( e^{-ik\cdot \hat{x}}F(\hat{x}) \right)
\end{equation}
where the Tr is defined as:
\begin{equation}
Tr\left(A\right) = Lim_{\Lambda \rightarrow \infty} 
\frac{(2\pi)^{2}}{\Lambda^{4}}\int^{\Lambda}d^{4}q<q|A|q>
\end{equation}
where the $|q>$ are momentum eigenkets.  This works because
\begin{equation}
<h|e^{-iq \cdot \hat{x}}e^{-ik\cdot \hat{x}}|h> = \delta (q - k)
\end{equation}

Given a second operator $G \Longleftrightarrow g$ , we have for the 
product:
\begin{equation}
FG \Longleftrightarrow f \star g
\end{equation}
where the star product is defined as:
\begin{eqnarray}
f \star g = Lim_{x^{\prime} \rightarrow x}e^{\frac{i}{2}\theta^{\alpha 
\beta}\partial_{\alpha}\partial_{\beta}^{\prime}}f(x)g(x^{\prime})\\
= f(x)g(x) + \frac{i}{2}\theta^{\alpha 
\beta}\partial_{\alpha}f(x)\partial_{\beta}g(x) 
\end{eqnarray}
to first order in $\theta$.

\section{The Snyder Star Product}

For a function f, using the Snyder operators $\hat{x}$ it is certainly possible 
to define a Hilbert space operator F as is done in Eqs.(25) and (26).  That is,
$f \Longrightarrow F$ .  However Eqs.(27) and (28) cannot be used to find f 
given F.  This is because the non-linearity of the momentum transformation shown 
in Eqs.(15) to (19) gives, instead of Eq.(29),
\begin{equation}
<h|e^{-iq \cdot \hat{x}}e^{-ik \cdot \hat{x}}|h> = \delta \left[hg(q,h) + 
qf(q,h) - hg(k,h) - kf(k,h)\right]
\end{equation}
and integrating this over all h yields an ordinary function of q and k, instead 
of the required dirac delta function.

Nevertheless, it is possible to go from F to f .  Let $|0>$ denote the zero 
eigenket of the momentum operator.  It can be shown that, to first order 
in $\theta$,
\begin{equation}
<0|e^{-iq \cdot \hat{x}}e^{-ik \cdot \hat{x}}|0> = \left(1 + 
\frac{\theta}{2}q^{2} \right)\delta(q - k)
\end{equation}
from which it follows that
\begin{equation}
\tilde{f}(k) = (2\pi)^{2}\left(1 - \frac{\theta}{2}k^{2} \right)
<0|e^{-ik \cdot \hat{x}}F(\hat{x})|0>
\end{equation}
and we also have $F \Longrightarrow f$ , for such functions F of $\hat{x}$ 
alone.

More generally, for any operator A, we could consider the function f(x) given by
\begin{equation}
f(x) = \int d^{4}k e^{ik \cdot x)}\left(1 - \frac{\theta}{2}k^{2} \right)
<0|e^{-ik \cdot \hat{x}}A|0>
\end{equation}

A problem arises when we consider products of functions of the 
$\hat{x}^{\alpha}$ , because the commutation rules (3) do not close on the 
$\hat{x}^{\alpha}$, but involve the $M^{\alpha \beta}$.  A product of functions F
of $\hat{x}^{\alpha}$ becomes a function of both the $\hat{x}^{\alpha}$ 
and the $M^{\alpha \beta}$ . Nevertheless, we can still use Eq.(36) to find a 
function of $x^{\alpha}$ associated with the operator product.   Let 
$f\Longrightarrow F$ and $g \Longrightarrow G$ .  Defining
\begin{equation}
\tilde{fg}(k) = (2\pi)^{2}\left(1 - \frac{\theta}{2}k^{2} \right)
<0|e^{-ik \cdot \hat{x}}F(\hat{x})G(\hat{x})|0>
\end{equation}
and $f \star g(x)$ as the Fourier transform of $\tilde{fg}(k)$, we get, to 
first order in $\theta$, the Snyder star product:
\begin{equation}
f \star g(x) = 
f(x)g(x) - Lim_{u,v \rightarrow x} \frac{\theta}{2}x \cdot 
\left[\nabla_{v} \left( \frac{\nabla_{u}^{2}}{2} + 
\nabla_{u} \cdot \nabla_{v} \right) - \nabla_{u} \left(\nabla_{v}^{2} 
  + \frac{\nabla_{u} \cdot \nabla_{v}}{2} \right) 
\right]f(u)g(v)
\end{equation}

The operators $\hat{x}^{\alpha}$ and $M^{\alpha \beta}$ generate an algebra 
acting on a Hilbert space H.  Eq.(36) is a map from operators on H to a 
Snyder $\star$ algebra of functions on Minkowski space.  It takes multiples of 
the operators $e^{\frac{i}{2} \omega_{\alpha \beta}M^{\alpha \beta}}$ into 
the identity since, to first order in $\theta$
\begin{equation}
e^{ih \cdot \hat{x}}e^{ik \cdot \hat{x}} = e^{ip \cdot 
\hat{x}}e^{\frac{-i}{2}h \cdot M \cdot k}
\end{equation}
where
\begin{equation}
p^{\alpha} = h^{\alpha} + k^{\alpha} + \frac{\theta}{3} \left[ k^{\alpha} 
\left( \frac{h^{2}}{2} + h \cdot k \right) - h^{\alpha} \left( k^{2} + 
\frac{h \cdot k}{2} \right) \right]
\end{equation}
and the state $|0>$, being invariant under Lorentz transformations, takes the 
operators $M^{\alpha \beta}$ into zero.

\section{Discussion}

The presence of a constant matrix $\theta^{\mu \nu}$ in the commutation 
relation Eq.(1) spoils not only Lorentz invariance, but also the isotropy 
of empty space-time.  The Snyder model, although it preserves Lorentz 
invariance, still violates spatial isotropy.  One sign of this is the different 
spectra for space and time coordinates.  Another is the presence of the 
coordinate x in Eq.(38), the Snyder star product.  This means the point x = 0 
is a preferred point, which is not surprising, as rotations and boosts 
leave this point invariant.
    
Eq.(36) gives a map from an algebra of operators on Hilbert space to the 
algebra of functions on coordinate space furnished with the Snyder $\star$ 
product.  The map cannot be an isomorphism, since some operators go into the 
identity.  It is not known if the map is some kind of homomorphism, in 
particular, it is not known if the Snyder $\star$ product is associative.

Is the Snyder algebra relevant?  This is a noncommutative geometry based 
on SO(4,1)  A similiar geometry based on SO(5) has recently been investigated 
in Ref.~\cite{grosse}


\begin{thebibliography}{9}
\bibitem{douglas}
Michael R. Douglas and Mikita A. Nekrasov, 
\emph{Noncommutative Field Theory}, 
hep-th/0106048.
\bibitem{sak}
Naoki Sasakura,
\emph{Space-time uncertainty relation and Lorentz 
invariance},
hep-th/0001161.
\bibitem{carroll}
Sean M. Carroll, Jeffrey A. Harvey, V. Alan Kostelecky, 
Charles D. Lane and Takemi Okamoto,
\emph{Noncommutative Field Theory and 
Lorentz Violation},
Phys Rev. Lett., {\textbf 87}, 141601 (2001), hep-th/0105082.
\bibitem{snyd}
Hartland S. Snyder,
\emph{Quantized Space-Time},
Phys. Rev., {\textbf 71}, 38 (1947).
\bibitem{grosse}
H. Grosse, C. Klimcik and P. Presnajder,
\emph{On Finite 4D Quantum Field Theory in Non-Commutative Geometry}, 
hep-th/9602115
\end{thebibliography}
\end{document}